# IoT- Based Low-Cost Soil Moisture and Soil Temperature Monitoring System

Guruprasad Deshpande[1]
guruprasad.deshpande19@pccoepune.org

Mangesh Goswami[2]
goswamimangesh71@gmail.com

Jayesh Kolhe[1]
jayesh.kolhe19@pccoepune.org

Vishal Khandagale[1]
vishal.khandagale19@pccoepune.org

Darshan Khope[1]
darshan.khope19@pccoepune.org

Gargi Patel[1]
gargi.patel19@pccoepune.org

Radhika Doijad[1]
radhika.doijad19@pccoepune.org

Rajani P.K.[1]
rajani.pk@pccoepune.org

Milind Mujumdar[2]
mujum@tropmet.res.in

Bhupendra Bahadur Singh[2]
bhupendra.cat@tropmet.res.in

Naresh Ganeshi[2]
naresh.ganeshi057@gmail.com

[1] Department of Electronics and Telecommunication, Pimpri Chinchwad College of Engineering Pune, India

[2] Centre for Climate Change Research, Indian Institute of Tropical Meteorology (IITM) (Ministry of Earth Sciences), Pune, India

***Abstract*—Soil moisture (SM) is referred to as a finite amount of water molecules within the pore spaces and it is a crucial parameter of Hydro-Meteorological processes. The behaviour of soil moisture water changes spatially and temporally in response to topography, soil characteristics, and climate[1]. Soil moisture is overseen by various hydro-meteorological factors that vary vertically with depth, laterally across terrestrial shapes, and temporarily in feedback to the climate. The precise monitoring and quantification of high-resolution surface and subsurface soil moisture observations are very important [13]. This paper highlights the outcomes of the fieldwork carried out at IITM, Pune, wherein we have developed a soil moisture and temperature measurement system using Raspberry Pi and the Internet of things (IoT). The development is classified into three stages, the first stage includes the assembly of the sensor with the microprocessor. The deployment of the low-cost system, data generation, and communication through a wireless sensor network is part of the second stage. Finally, the third stage includes real-time data visualization using a mobile application and data server for analysing soil moisture and temperature. The soil moisture profile obtained through the sensor deployed is highly correlated (r=.9) with in-situ gravimetric observations, having root means square error (RMSE) of about 3.1%. Similarly, the temperature observations are well-matched with the in-situ standard temperature observation. Here we present the preliminary results and compare the accuracy with the state-of-the-art sensors.**

**Keywords:** Soil Moisture, Soil Temperature, Raspberry Pi, Internet of things (IoT), Things Speak Platform IITM- COSMOS site

## I. Introduction

Soil moisture (SM) variability is a crucial component of agricultural, natural hazards (landslides and debris flow), ecological, hydro-meteorological, and climate studies. The availability of SM data helps deduce important conclusions in these fields. In operational hydrology, SM is one of the important state variables [2]. *S*oil water content varies greatly over time and across regional scales. From centimetres to decimetres, many parameters such as soil texture, vegetation distribution, bulk density, precipitation, and irrigation vary [3]. The soil layer is split into four layers based on the standard deviation and coefficient of variation (CV) of SM: the quick-change layer, active layer, sub-active layer, and generally stable layer. [4]. At 30 cm, the distinction between the active and sub–active layers is made. The difference between rainfall and evapotranspiration has been found to explain a major part of the dependability characteristics of SM content in forest ecosystems. Evapotranspiration is a key factor that affects the stormwater retention capacity of green roofs, and consequently their hydrological performance [5][6]. Due to the variability of environmental elements such as water, energy input, and soil texture, moisture changes vertically in-depth and laterally. Due to various applications in the different research domains., accurate and precise measurement of SM is required. Various methods have been developed to record and analyze it.

Measurement of SM is done through various techniques such as remote sensing, and in-situ-based platforms. Classical SM measurement methods like a thermo-gravimetric, sensors-based approach like these are classified into three categories like resistive sensor, conductive sensor, inductive sensor, dielectric methods, electromagnetic induction, tensiometer method, neutron moisture meter, etc [8]. The basic method for measuring SM is the gravimetric method, which is regarded as the conventional and accurate method adopted for measuring the quality of SM measurements which are based on other techniques. It is also used for calibration purposes. Measurement of SM using a capacitive SM sensor is popular due to its low cost and better accuracy. This sensor can be used for the measurement of SM at different depths. These measurements can be standardized with the help of calibration through the gravimetric method [9][10]. Continuous SM measurements at specific intervals and analysis are required to prepare hydrological models for climate modelling. Modern techniques like IoT and cloud storage can be used to perform this analysis. Many microprocessors and microcontrollers are used in this method for this purpose, like Arduino [11], where a data logger is used. In this approach, the cost of the system design and complexity causes scalability issues.[12]

In this paper, we have developed a network of low-cost SM conductive sensors and temperature sensors using Raspberry Pi and IoT. The optimization of low-cost SM and temperature profile monitoring systems evolved at the COSMOS-IITM site [13] is a part of the CCCR-IITM internship project work. This development comprises low-cost capacitive sensor SM and temperature sensors. This system measures soil parameters at different depths across one profile. This system can be classified into three stages; the first stage includes the designing and assembling of a sensor deployment system. The second stage includes the designing and implementation of a wireless sensor network for the observation of SM and temperature profiles, and the third stage includes real-time data visualization/analysis of SM and temperature of the field scale network for modelling using various software. Additionally, we have designed a web portal for better user experience and CSV format of data can be generated through this system to analyse data. This is a fully automatic system. Further, this paper describes the methodology and safety measures taken while assembling and developing the system along with the calibration and the standardization of the values received.

## II. METHODOLOGY

As the system is sending data using IoT with the help of sensors collecting the data from the soil. The following diagram illustrates the generalized work of the system with a set of blocks used in the system.

The block diagram systematically describes the system. The Raspberry Pi is used as a microprocessor to process the data collected from sensors. The sensors like SM measurement sensors and temperature sensors are used in the system. The network of 4 pairs of SM and temperature sensors is made to measure the parameter of different soil levels inside the ground. These sensors are connected to the Raspberry Pi's GPIO, and an analog to digital converter is used to connect an SM measurement sensor to the Pi. The system is developed with the help of a low-cost sensor, it sends the data over an open-source cloud thing speak using the MQTT protocol, The data collected by these sensors are also displayed on the website also the CSV file is generated of this data to analyse and calibrate the sensor

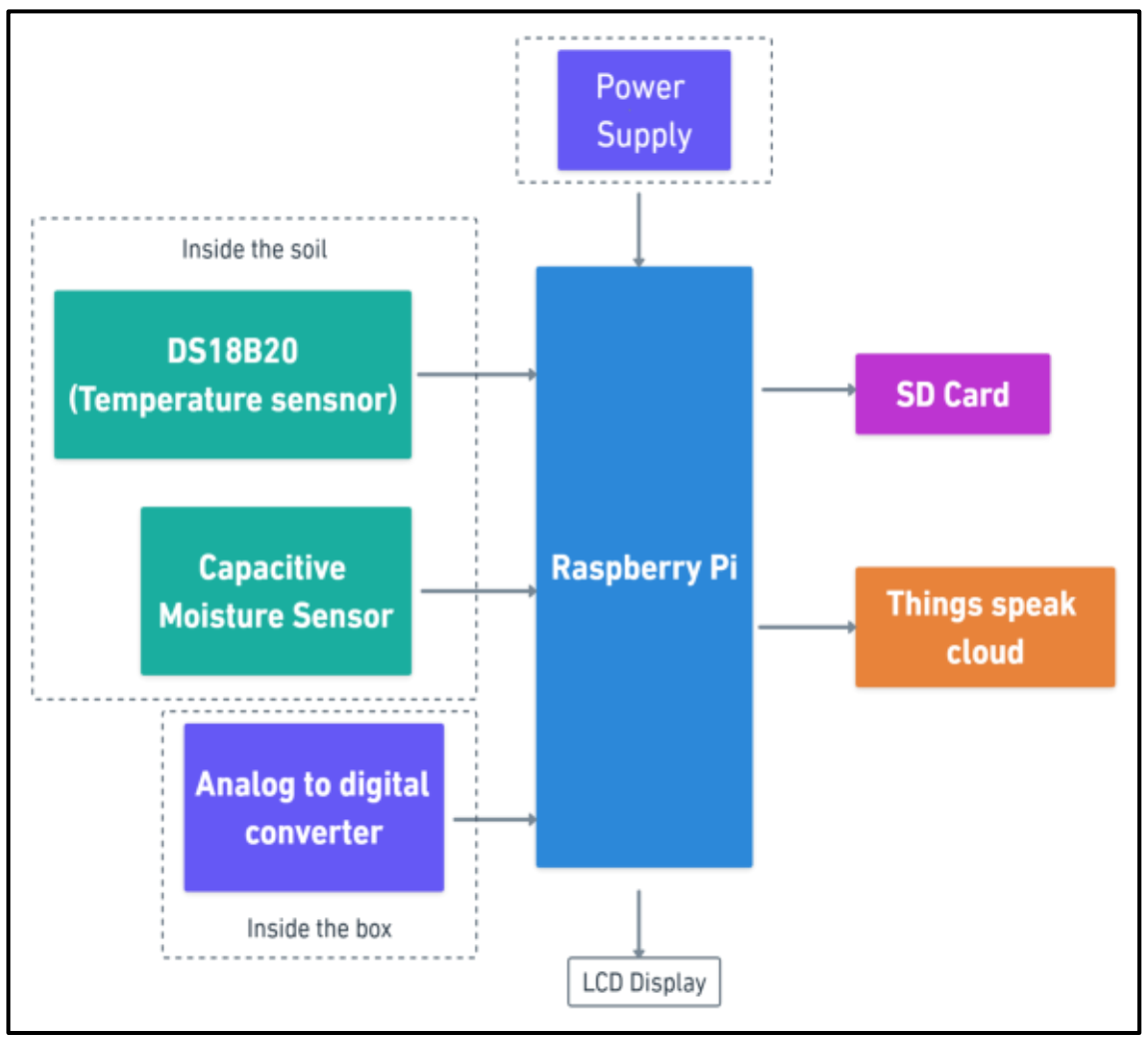

**Fig. 1.** Block Diagram

As described in the block diagram **Fig. 1** Raspberry Pi is used as a microprocessor. It is a 40-pin board having different pins for different purposes like power pins, ground, etc, and the 28 pins are used as general-purpose input-output for interfacing different sensors which means we can program these pins according to our use of the sensor.it is supported by Raspbian OS which is similar to Linux, therefore, it is easy to handle and analyse. this processor can be connected with a cable to connect it to the internet. It also supports the wireless connection of the WI-FI network. These features make raspberry pi an ideal processor for applications like monitoring systems and IoT.

Temperature is measured using the temperature sensor DS18B20 temperature sensor. In the system, a probe sensor with waterproof shielding is used. It communicates with an inner microprocessor using the Dallas one-wire bus protocol, which employs only one data line. The DS18B20 provides temperature readings in the 9-bit to 12-bit range. The voltage across the diode terminals is the primary working mechanism of temperature sensors. When the voltage rises, the temperature rises as well, resulting in a voltage drop between the base and emitter transistor terminals of a diode. It directly converts temperature into a digital value. This sensor does not require an analog to digital converter as it has a digital output.

To measure SM there are various methods. In this system, the dielectric method is used to measure the moisture of the soil as it is a low-cost sensor method. The capacitive SM sensors are used in the system and soil acts as the dielectric. This capacitive moisture sensor operates by detecting variations in capacitance produced by soil dielectric changes. The dielectric created by the soil is measured capacitively, and the most essential factor that impacts the dielectric is water. this analog sensor is low cost and good for applications like soil monitoring. To get appropriate reading this sensor needs to be calibrated with standard sensors.

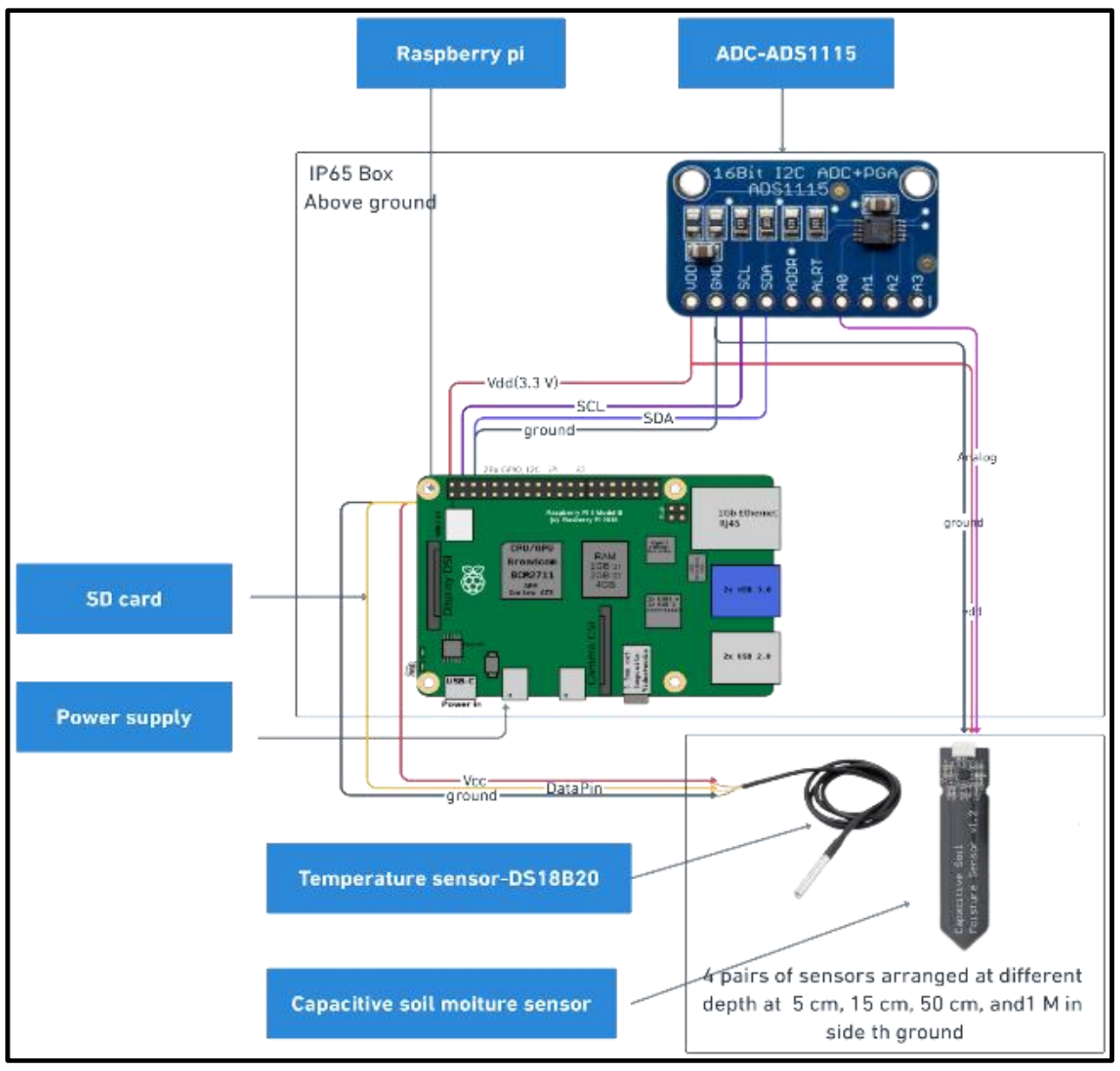

**Fig. 2.** Circuit Diagram

As the capacitive sensor is analog out, it has been converted into digital output using 16-bit Analog to digital converter (ADC) -ADS1115 which gives a digital value of the analog sensor up to 4 decimals after a point. Also, to connect various sensors this ADC is useful. As the sensor is capacitive the soil and moisture content present in it acts as a dielectric material for the capacitive sensor. The capacitance is inversely proportional to voltage. this shows that the output voltage decreases when the moisture content is increased and vice versa. It has various pins like SCL (serial clock) SDA (Serial Data) and 4 analog out pins. The ADC sends data over two lines using the I2C interface. Raspbian is the operating system that the Raspberry Pi runs on. This is a highly optimised operating system. It's for the Raspberry Pi range of ARM-based small single-board computers. All programs have been running on this OS along with real-time and other computer processes. The code is put in the start-up of Raspberry Pi i.e., in the bashrc of the OS. This ensures that the program is automatically started at the start-up of the Raspberry Pi board. This makes the system autonomous and reliable.

IoT is used for data transmission and data storage. Data from Raspberry Pi is directly sent to the thing speak cloud, thing speak saves all the data and is used for further work. It is attached to the website. The website is made by PHP used for the backend of the website, the gateway of sign-in and signup system to the database is done with the help of this tool. HTML, CSS, JS these languages used for the frontend of the website, website design is done with help of these languages. MySQL database is used to store data. The website has features of hashing security for passwords so that unauthorized users can't access or decrypt the password, the complete responsive architecture is supported for mobile, tablet, and laptop screens. Auto importing data from thing speak cloud without refreshing the browser. After data is sensed by Raspberry Pi, data is sent on the thing speak cloud. All the data is saved on the thing speak cloud, which can be used for further purposes. That data is imported into the website and then monitored on the website for further evaluation. This system is programmed mainly in python programming language as Raspberry Pi supports python, with the help of different libraries like urllib i.e., used for making web requests, Date-Time, time i.e., used to give delays and other time functionalities, etc. the functionalities are given in the project. In fact, these invasive soil sensors need settlement time for obtaining stable output signals.

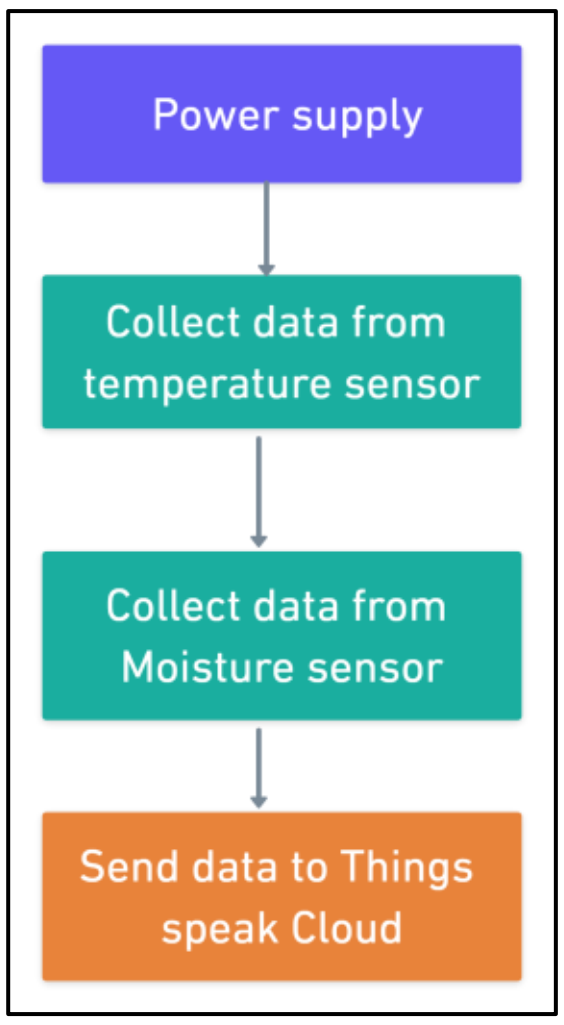

**Fig. 3**. Flow chart

The raw data from the DS18B20 sensor is acquired using one wire connection. This sensor works on Dallas's one-wire bus protocol which enables the connection of multiple sensors to the same pin as this is a bus protocol. The system uses 4 temperature sensors. Each sensor has its address which when called by the processor communicates the data collected. This information is saved in the Raspberry Pi's directory, which is accessed through the Integrated Development Environment (IDE), and the raw data is processed using Python programming. The temperature data was then calculated and sent to the Things Speak cloud using a communication protocol called Message Queuing Transport of Telemetry (MQTT). To support communication between IoT devices, this protocol employs a publish-subscribe method. MQTT allows client devices to publish messages to a central broker, which then allows other client devices to subscribe to the messages, allowing the broker to mediate communication between the publisher and the subscriber. Things speak has added an MQTT broker so that devices can send messages to things speak.

The analog data from the capacitive moisture sensor is collected by the ADC. Raspberry Pi collects this digital information. The voltage information is saved in a file. The temperature sensors' data is provided to Thing speak in the same way as the temperature sensors' data. The final code comprising both of these actions simultaneously is done by putting the program in the start-up of the Raspberry Pi. This ensures that the program is run whenever the Raspberry Pi is powered up and starts continuing the data. The finished software uses both sensors at the same time. The same procedures as before are followed, but with a new function. The information gathered by the various functions is saved locally in CSV format. The data from both sensors are shared on platforms such as Think Speak Cloud.

The whole system is deployed onto the site by taking various safety measures to get a correct reading with minimum loss and accurate measurement of intended parameters. With help of hardware and software, this system is made autonomous. The system starts and gives the data over the internet whenever powered up at an interval of 15 minutes In this study, these sensors are deployed 4 pairs of SM and temperature sensors, placed at 1-meter intervals on four different soil levels. The sensors are connected to a Raspberry Pi that produces output on our screens. It has been programmed in such a way that we can watch the output.

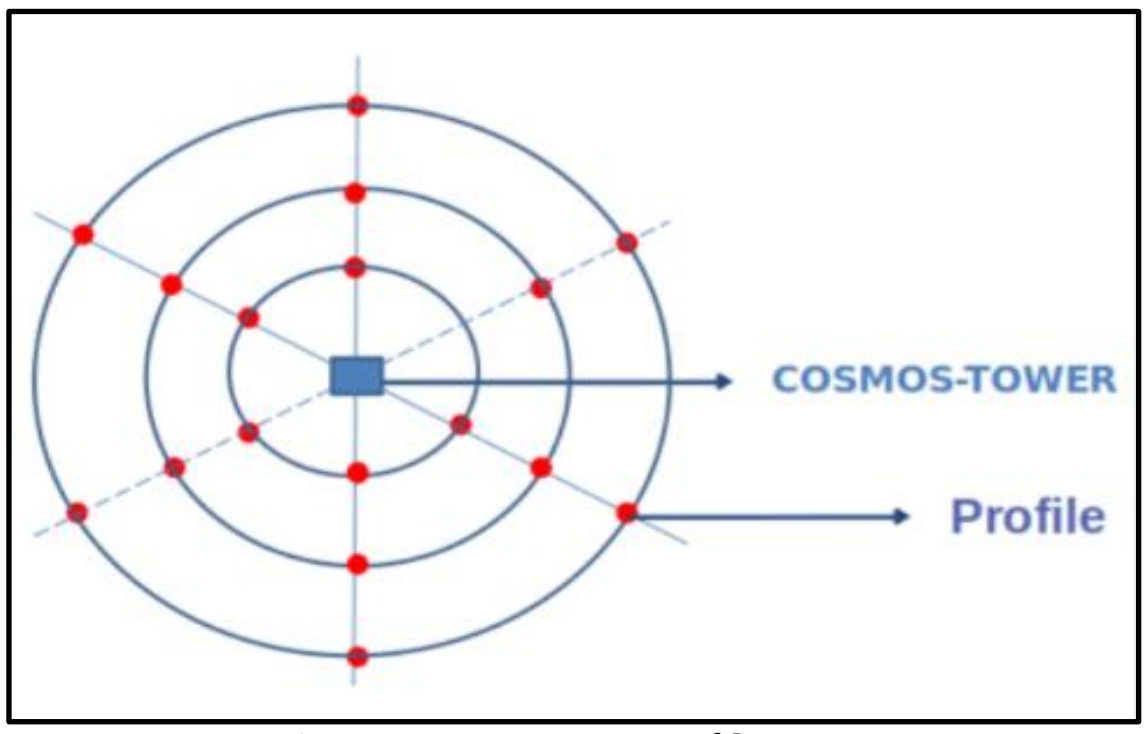

**Fig. 4.** COSMOS site profile points

In the above figure 4, the COSMOS Tower is at the centre which is denoted by blue-coloured squares, and profiles are denoted by the red-coloured circles at specified distances from the COSMOS-TOWER [13]. Each profile has a one-meter depth and it contains four SM sensors and four temperature probes. One profile is shown in the following figure 4: example profile. These eight sensors were installed in each profile in the huge land area at specific distances as shown in figure. The first SM sensor is installed from the top at a 5 cm distance, the second one is at 15 cm, the third one is at 50 cm, and the last fourth SM sensor is at 100 cm. The Raspberry Pi is above the ground in the box. These four SM sensors are connected to the Raspberry Pi through which data is sent to the server. The server receives the collected data from the sensors in comma-separated values i.e., CSV file format.

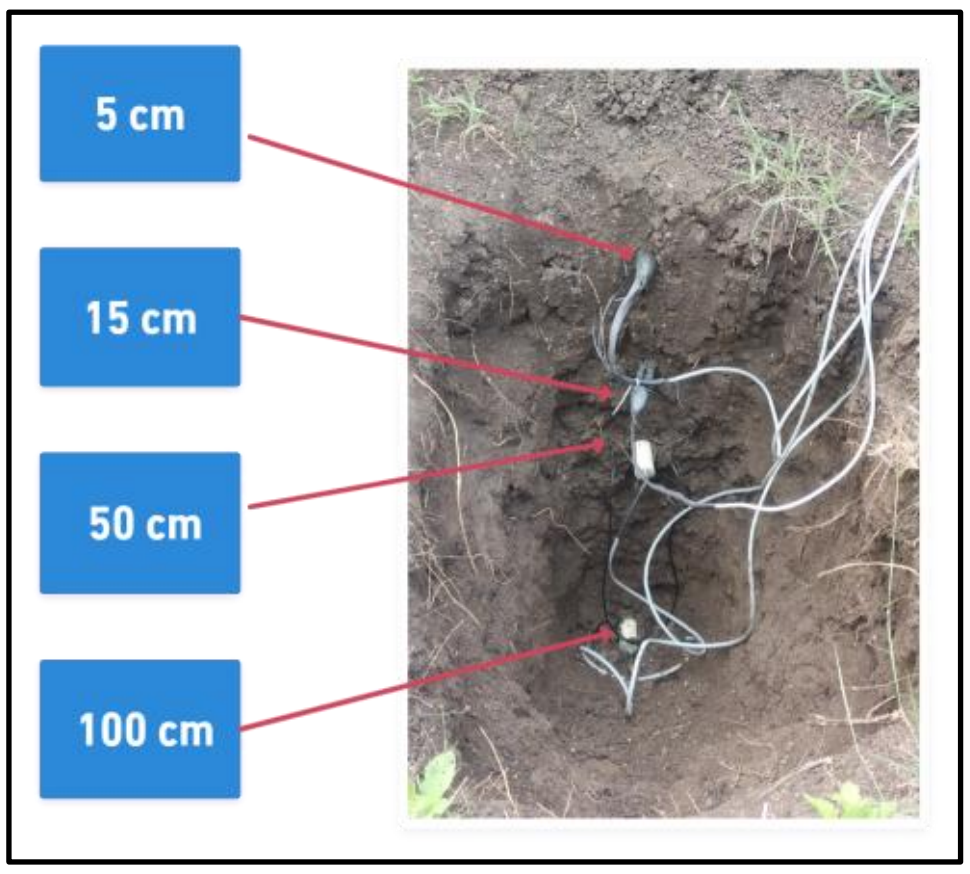

**Fig. 5(a)** Deployment of sensor at various levels in the depth

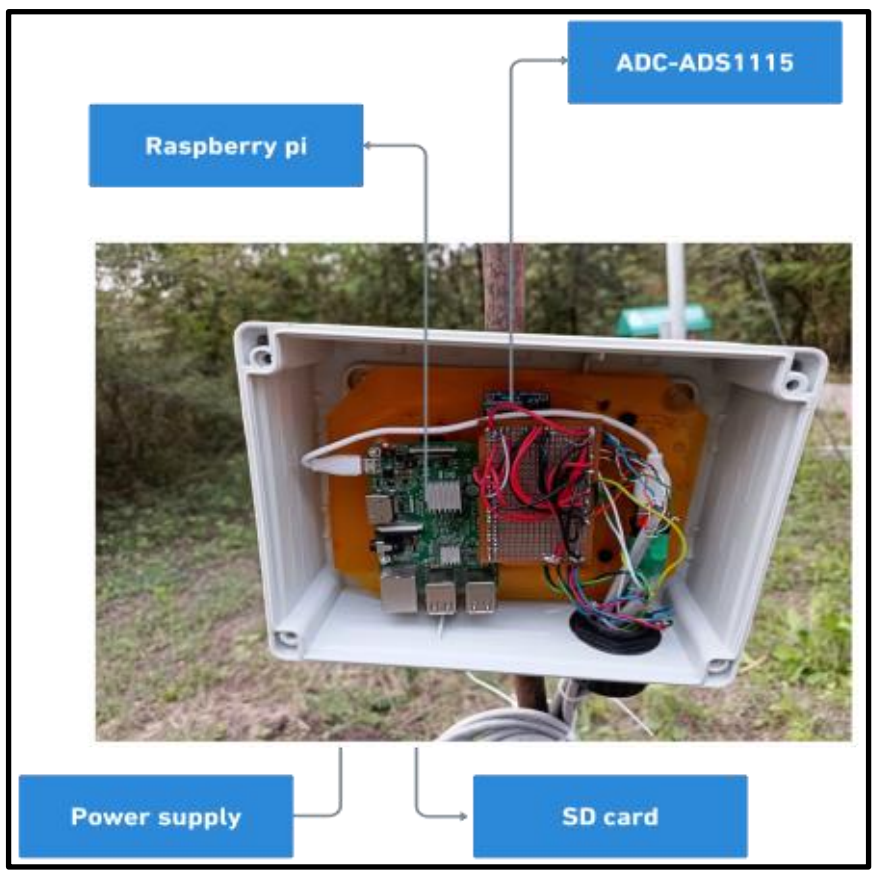

**Fig. 5(b)** Deployment of Hardware System.

The system communicates data using CAT 6 cable, it is used for connecting sensors to the Raspberry Pi and for the power supply. As it has a bandwidth of up to 250 MHz and a speed of up to 10 Gbps, so this cable helps for less data loss for the large distance. Its outer sheath is made up of strong material so it can't be harmed by any insects, soil, or moisture that can't reach the inner copper wires. The kind of care is taken while working with the sensors. Heat shrink/ insulation is used to insulate the sensors for the desired part for preventing them from damaging by any type of insects, soil or to protect them from SM/water to reach the sensitive part of the sensors. A heat sink is used in Raspberry Pi in order to cool it done efficiently as it is used 24/7. The heat sink takes away the increase in heat flow from the device by increasing the amount of low-temperature fluid across the surface. They used an IP66 box that is composed of polypropylene, reinforced glass fibre, thermoplastic elastomer sealing, and 10 $mm^2$ terminals/poles. With a Quick-Release Lid and Plastic Screws that close the box with a quarter turn, it is captive and rust-free. Also, it is weatherproof, ultraviolet resistant, shatterproof, and impact resistant.

## Calibration:

The capacitive sensors are deployed at the IITM- COSMOS site [11] and the data collected from the sensors are sent a Mobile application through IoT every 15 min. This can be observed on things speak IoT platforms **Fig. 6** this data can be accessed in various formats like CSV, XML, and JSON.

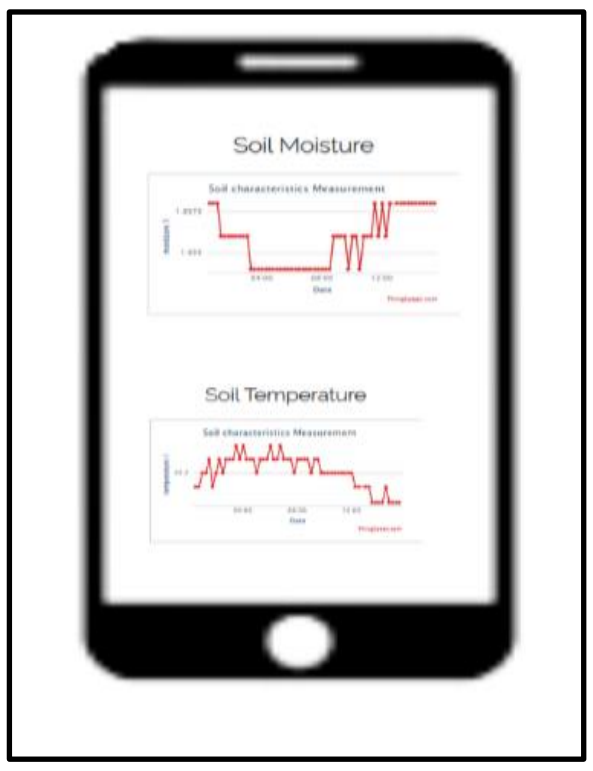

**Fig. 6(a).** Soil moisture and temperature data representation on the Mobile application and website

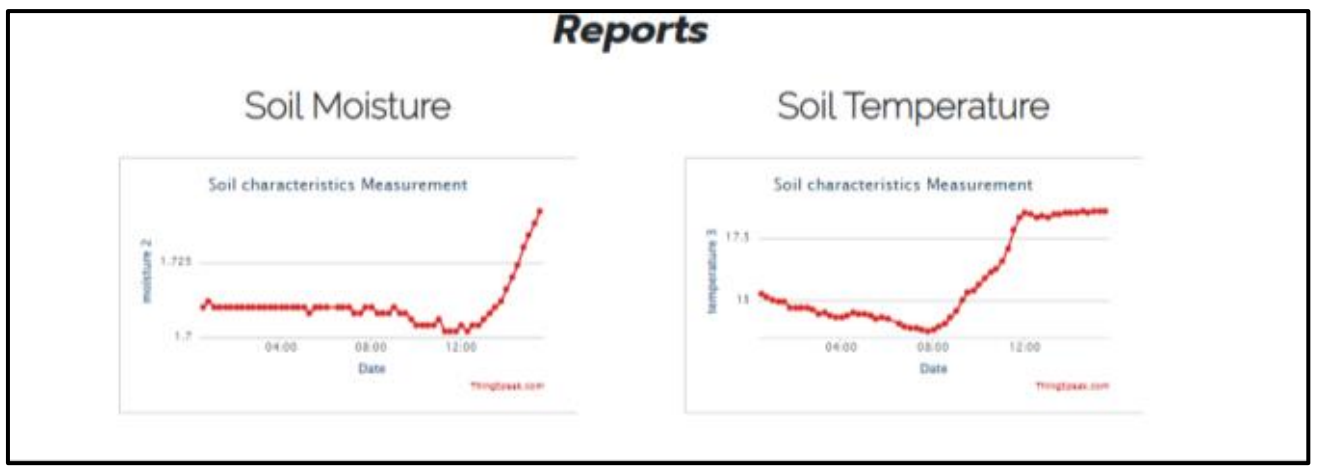

**Fig. 6(b).** Soil moisture and temperature data representation on the website

Simultaneously, soil samples from various locations and depths were collected and processed using the conventional gravimetric soil analysis method at the IITM-COSMOS site. The samples were kept in a plastic bag to prevent SM loss during the calibration, and they were then sealed with a lid and stored. The weight of the soil samples was obtained first, and then the samples were baked in an oven at 100 degrees Celsius for 24 hours, after which the weight of the soil samples was taken again, and the volumetric water content was estimated. The ratio of the volume of water to the volume of soil is the volumetric water content ($\theta_v$) of a sample [7]. The technique described above is repeated for different locations and the soil types procedure is repeated for different locations and different soil types.

$$\theta_v = \left(\frac{m_s - m_d}{m_d}\right)\frac{\rho_{d,s}}{\rho_w} \qquad (1)$$

where,

$m_s$ - a measured mass of the wet soil

$m_d$ -measured mass of the dry soil

□$_{d,\ s}$ - bulk density of soil (mass of dry soil divided by sample volume)

□$_w$ - density of water

The measured data from the Gravimetric sensor ($\boldsymbol{\theta_V}$) and observed sensor output data are compared and analysed as per the standard procedure. It is observed that Volumetric water content ($\boldsymbol{\theta_{Vnew}}$) is computed following the best fit [14] carried out through rigorous calibrations and given as:

$$\theta_{Vnew} = (-71.789x^2 + 158.04\,x - 37.711) \qquad (2)$$

Here **x** is the sensor output voltage.

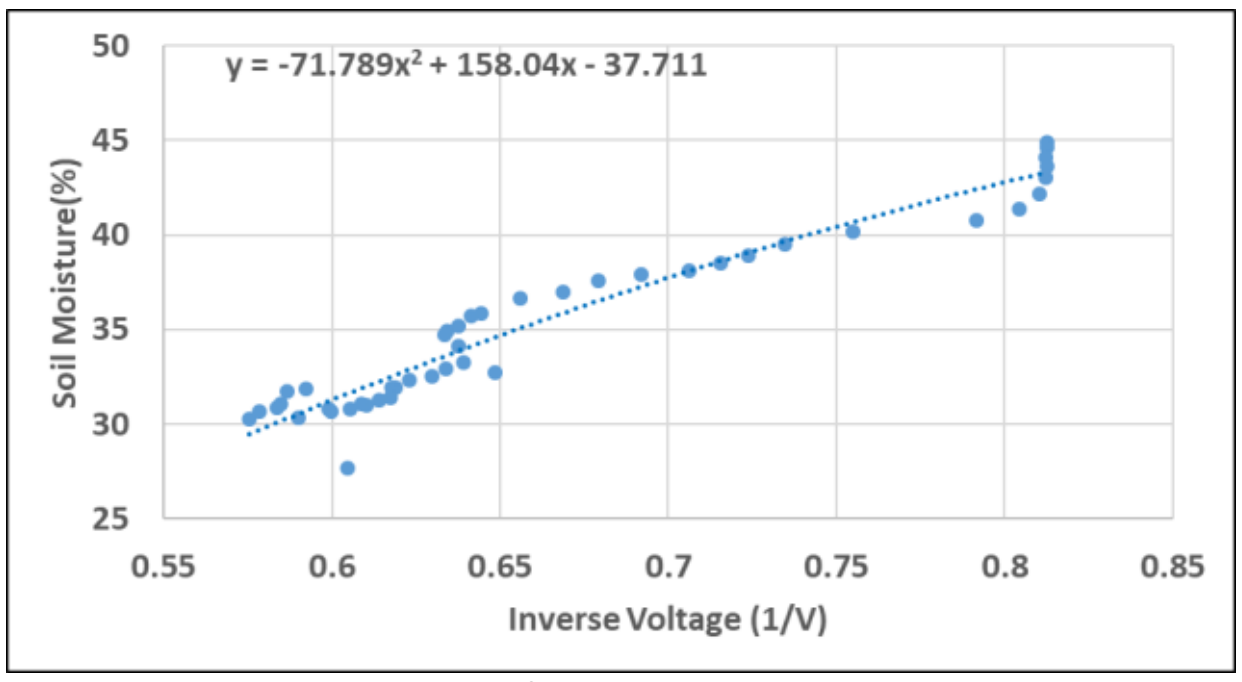

**Fig. 7.** shows the results of the calibration procedure and the relation between the inverse of the voltage reading from the capacitive sensor and θv derived by gravimetric methods. The inverse of the voltage reading was fitted via least-squares regression, which resulted in the relationship between the capacitive sensor and θv (Equation 2).

## III. Results and Analysis

These SM and temperature profile data sets obtained using the assembled low-cost observation system at the COSMOS-IITM site for 5 cm, 15 cm, 50 cm, and 100 cm depth are validated and analysed to explore surface and subsurface variations. The techno-scientific expertise, distinctive infrastructural facilities, and ample in-situ hydro-meteorological data sets at the COSMOS-IITM site having natural vegetation cover [13] have extended valuable support for calibration and validation in this study. The correlation coefficient of SM profile observations with those of in-situ Gravimetric data sets is about 0.8, however, the Root Mean Square Error (RMSE) is found to be about 3.1 %. The temperature profile observations match very well (~95%) with those obtained from standard sensor-based temperature measurements. Figure 8 (a and b) shows the time series of SM and temperature observations at 5 cm, 15 cm, 50 cm, and 100 cm depth. The variations at surface levels (5 and 15 cm) are relatively higher than those of subsurface levels (50 cm and 100 cm). The distinction between SM and temperature variations at surface and subsurface are remarkable. However, the soil moisture is higher in the subsurface levels and persists smoothly. While the soil temperature was found to be lower in subsurface levels.

The data observed for a period of about 2 months. Analysis of that data is shown in Table I, Table II and Table III respectively. The analysis of the data received form soil moisture and temperature sensor as discussed in above result and analysis..

**TABLE I**. RMSE AND CORRELATION WITH RESPECT TO GRAVIMETRIC OBSERVATIONS

| DATA SET | RMSE | CORRELATION |
|---|---|---|
| LOW COST SENSOR | 0.03 | 0.90 |
| STANDERD SENSOR | 0.02 | 0.94 |
| GLDAS | 0.05 | 0.87 |
| ERA | 0.07 | 0.85 |

**TABLE II.** RAW ANALOG VALUES(VOLT) AND PROCESSED OUTPUT (VOLUMETRIC WATER CONTENT)

| RAW ANALOG VOLTAGE (VOLT) | PROCESSED VWC IN % |
|---|---|
| 1.23 | 43.21 |
| 1.24 | 42.96 |
| 1.26 | 42.40 |
| 1.32 | 40.68 |
| 1.36 | 39.65 |
| 1.38 | 39.07 |
| 1.40 | 38.62 |
| 1.42 | 38.09 |
| 1.45 | 37.28 |

**TABLE III.** MIN AND MAX VALUES OVER THE STUDY PERIOD

| | |
|---|---|
| MINIMUM VWC (%) | 29.46 |
| MAXIMUM VWC (%) | 43.31 |
| MINIMUM TEMP (DEG C) | 14.98 |
| MAXIMUM TEMP (DEG C) | 23.79 |

The plots for the data collected are shown in the following fig 8 (a) and fig 8 (b) for SM and temperature

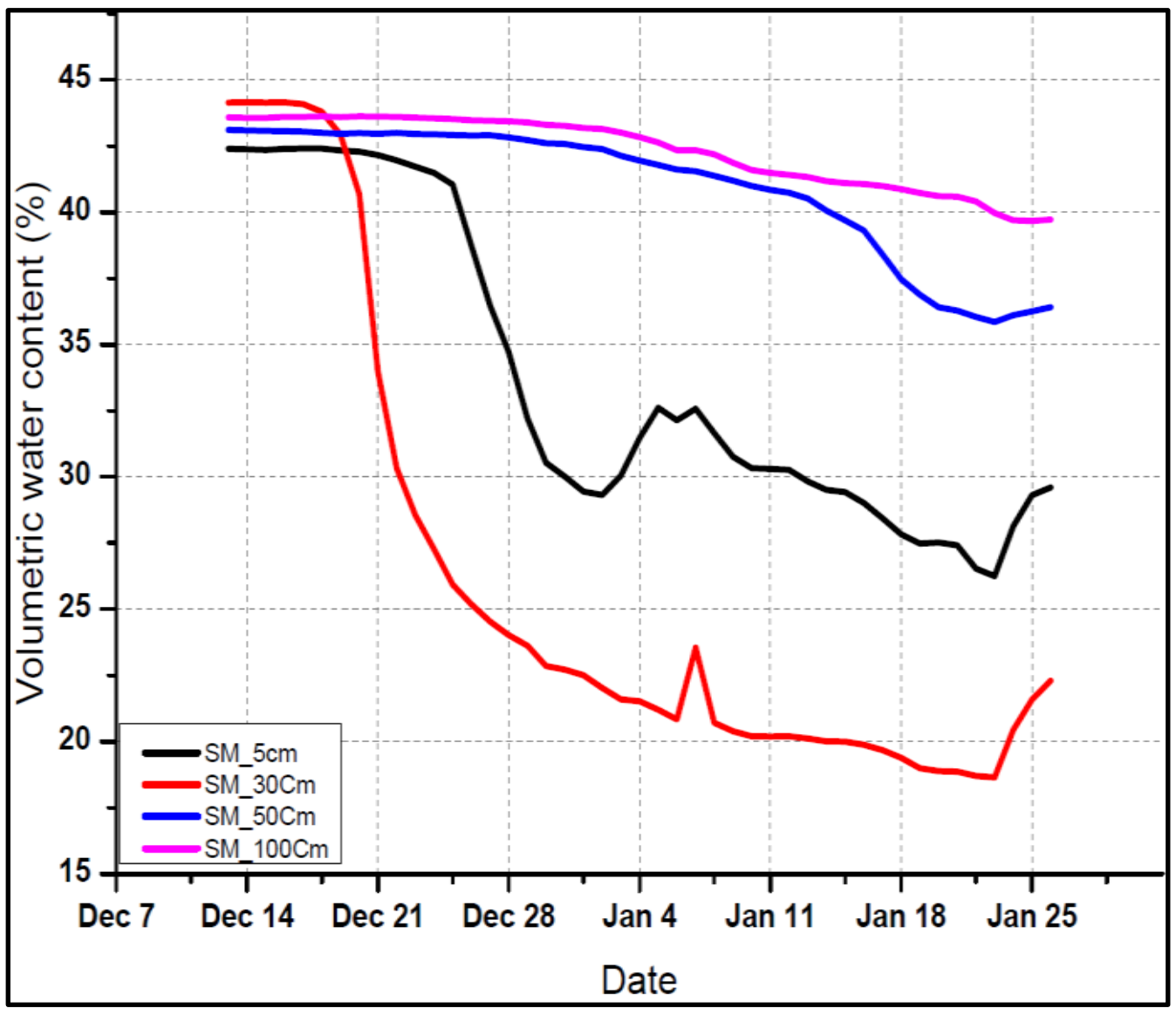

**Fig. 8 (a).** Soil moisture variation

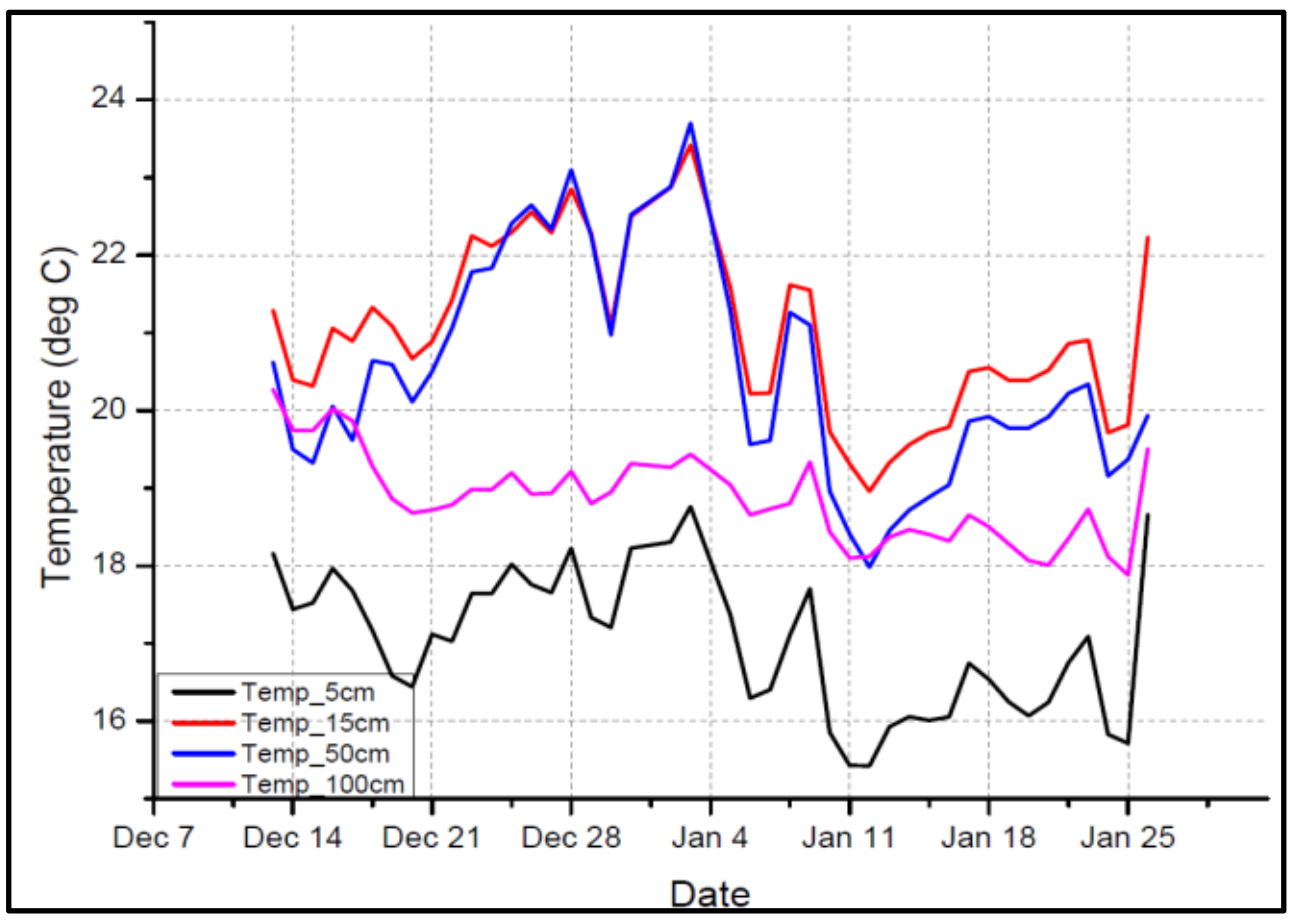

**Fig. 8(b).** Soil Temperature variation

## IV. CONCLUSION

The optimization of low-cost SM and temperature profile monitoring systems evolved at the COSMOS-IITM site [13] is a part of the CCCR-IITM internship project work. A capacitive sensor and temperature sensor are used here to monitor SM and temperature at surface (5 and 15 cm) and subsurface (50 and 100 cm) levels. This system has been developed in three steps including assembly design, data management, and real-time visualization with analysis. The credentials of SM and temperature observations are set through critical calibration and validation using standard in-situ data sets available at the COSMOS-IITM site. Interestingly, real-time data access through mobile using things speak cloud or data server is a crucial component of this system. The time-series analyses depict the higher variability of surface SM with lower magnitudes as those of higher magnitude of subsurface SM. Also, the distinct lower subsurface temperature variations, having lower magnitude, compliments the subsurface SM variations. The real-time knowledge of the variation of SM and temperature profiles is crucial for hydro-meteorological, agricultural modelling, and various applications.

Researchers can study different crops and moisture variations to optimize the irrigation practices with this observation system. Also, SM sensors can be customized to fit a variety of soil types. Bluetooth, LoRA can be used to provide the output data straight to the user wirelessly without internet connectivity according to the application. A GPS module can be fitted to determine the amount of moisture in a specific area. A camera can also be added to this observation system to periodically record sky and ground conditions to monitor crops and their health.

## ACKNOWLEDGMENT

The authors are grateful to the Director, Indian Institute of Tropical Meteorology (IITM, India), and the Director Pimpri Chinchwad college of Engineering (PCCOE) Pune. for their unconditional support to carry out this research work at IITM. The logistic support provided by the IITM for maintaining this COSMOS-IITM observational site is duly acknowledged.

## DATA AVAILABILITY STATEMENT

The gravimetric observations and the observed profile data from the capacitive sensors at IITM-COSMOS site, can be made available on request with proper permission form The Director IITM and Director PCCOE.